# Closed-Loop MOEMS Accelerometer

**MAJID TAGHAVI[1], ABOLFAZL ABEDI[1], GHOLAM-MOHAMMAD PARSANASAB[1,*], MOJTABA RAHIMI[2], MOHAMMAD NOORI[1], HAMZEH NOUROLAHI[1], HAMID LATIFI[3]**

[1]Integrated Photonics Laboratory, Faculty of Electrical Engineering, Shahid Beheshti University, Evin, Tehran, 1983963113, Iran.
[2]Department of physics, University of Isfahan, Isfahan, Iran
*Faculty of Electrical Engineering, Shahid Beheshti University, Evin, Tehran 19839-63113, Iran
[3] Faculty of Physics, Shahid Beheshti University, Tehran 19839-63113, Iran
*gm_parsanasab@sbu.ac.ir

**Abstract:** In this paper, a closed-loop micro-opto-electro-mechanical system (MOEMS) accelerometer based on the Fabry-Pérot (FP) interferometer is presented. The FP cavity is formed between the end of a cleaved single-mode optical fiber and the cross-section of a proof mass (PM) which is suspended by four U-shaped springs. The applied acceleration tends to move the PM in the opposite direction. The arrays of fixed and movable comb fingers produce an electrostatic force which keeps the PM in its resting position. The voltage that can provide this electrostatic force is considered as the output of the sensor. Using a closed-loop detection method it is possible to increase the measurement range without losing the resolution. The proposed sensor is fabricated on a silicon-on-insulator wafer using the bulk micromachining method. The results of the sensor characterization show that the accelerometer has a linear response in the range of ±5 g. In the closed-loop mode, the sensitivity and bias instability of the sensor are 1.16 V/g and 40 μg, respectively.

## 1. Introduction

Due to their small size, low cost, and reliable performance, MEMS accelerometers are employed in various fields such as vibration measurement, inertial navigation, and automotive applications [1–5]. The mechanical structure of a MEMS accelerometer has a suspended proof mass (PM) which moves when an acceleration is applied to the sensor. The working principle of these sensors is the detection of the mechanical displacement due to the applied acceleration. There are different approaches to measure mechanical displacement such as the capacitance, piezoresistive, and optical methods[6–10]. In optical MEMS (MOEMS) accelerometers, the PM displacement changes the characteristics of the output light. This optical change can be used to determine the magnitude and direction of the applied acceleration. Intensity and frequency modulations are two main approaches for measuring acceleration in MOEMS accelerometers [11–13].

In the first method, acceleration changes the intensity of the output light. Although this method is simple and low cost, it cannot provide a high sensitivity. The frequency modulation method is more accurate and has a higher bandwidth [14, 15]. This method is based on the phase shift in the output optical spectrum of the sensor. The photonic band gap (PBG) effect and interferometry are the most common approaches to implement the frequency modulation method [16–18]. The structure of most PBG-based sensors consists of nano-scale photonic crystals which have a sophisticated fabrication procedure. Among interferometry-based sensors, Fabry-Pérot (FP)-based MOEMS accelerometers have simpler configuration and an easier fabrication process. In FP-based sensors, acceleration changes the length of the FP cavity. This leads to an optical spectrum shift in the sensor output by which the applied acceleration can be measured[19–21]

The optical sensitivity and the measurement range are two principal characteristics of the sensor in FP-based MOEMS accelerometers. The maximum achievable spectral shift is limited to the free spectral range (FSR) of the FP cavity. Hence, there is a limitation in achieving a wide measurement range and a high sensitivity simultaneously [22–26]. To overcome this problem, a closed-loop MOEMS accelerometer is proposed in this paper. The closed-loop detection technique is common in MEMS accelerometers [27–29]. Moreover, it has recently been implemented in grating-based accelerometers [30–32]. In closed-loop MOEMS accelerometers, an external force (here an electrostatic force) is applied to the structure. This external force compensates the mechanical displacement of the PM due to the applied acceleration. This electrostatic force is used to measure the magnitude and direction of the applied acceleration. Because ideally there is no mechanical displacement in this accelerometer, there is no mechanical limitation to achieve a wide measurement range and a high sensitivity at the same time.

The organization of this paper is as follows. In section 2, the working principle of the device is discussed. In addition, the design and simulation of the optical, mechanical, and electrical parts of the sensor are described. In section 3, the fabrication process as well as the results of the sensor characterization in both open-loop and closed-loop modes are presented. Finally, the conclusions are given in section 4.

## 2. Design and Simulation

### 2.1 Working Principle

The closed-loop MOEMS accelerometer has two main systems, i.e. the sensing system and the control system. The sensing system is based on the Fabry–Pérot (FP) interferometer. The FP cavity is formed between the cross-section of the PM and the end of the optical fiber. As shown in Fig. 1-a, the PM is suspended by four U-shaped springs. Furthermore, an array of capacitors is formed between the fixed and movable comb fingers. The fixed and movable comb fingers are attached to the anchors and the PM, respectively. They are used to apply the external electrostatic force. This electrostatic force compensates the displacement due to the acceleration and prevents the movement of the PM.

In the open-loop mode, acceleration leads to a change in the length of the FP cavity. This change causes a shift in the output optical spectrum. As shown in Fig. 1-b, when the length of the cavity increases, the spectrum moves to longer wavelengths (red shift) and vice versa (blue shift). This movement can be detected using three different approaches, i.e. spectral monitoring, intensity monitoring, and cavity resonance method. In the first method, the spectrum shift can be measured using an optical spectrometer. In the second method, the intensity variations of a specific wavelength are monitored using a photodetector. The third method is based on lock a laser to the cavity resonance and measuring the optical frequency. Measuring the spectrum requires high resolutions spectrometer and is not cost-effective to build a commercial sensor. The second method (i.e., using a fixed laser wavelength) is cost-effective, compact, and easy to set up. Although the third method is precise, it is more complicated in comparison with other methods. Therefore, we use the second method which is more practical and affordable method.

As mentioned in the previous section, the measurement range and sensitivity of the sensor are interdependent due to the repetitive pattern of the FP optical spectrum. Therefore, in open-loop sensors, a wide measurement range leads to a low sensitivity and vice versa. In the closed-loop mode, there is almost no mechanical displacement and spectral shift due to the applied electrostatic force. Hence, having a wide measurement range and a high sensitivity simultaneously is achievable.

In this method, the applied acceleration tends to move the PM. This movement leads to intensity variations in the photodetector. The output voltage signal of the photodetector is sent to the control system. The control system calculates the required electrostatic force and the appropriate voltages. Then, the calculated voltage is applied to the electrodes. The electrostatic force returns the PM to its resting position and the control loop is completed. The calculated voltage in the control system is considered as the output of the accelerometer. This voltage determines the amplitude and direction of the applied acceleration.

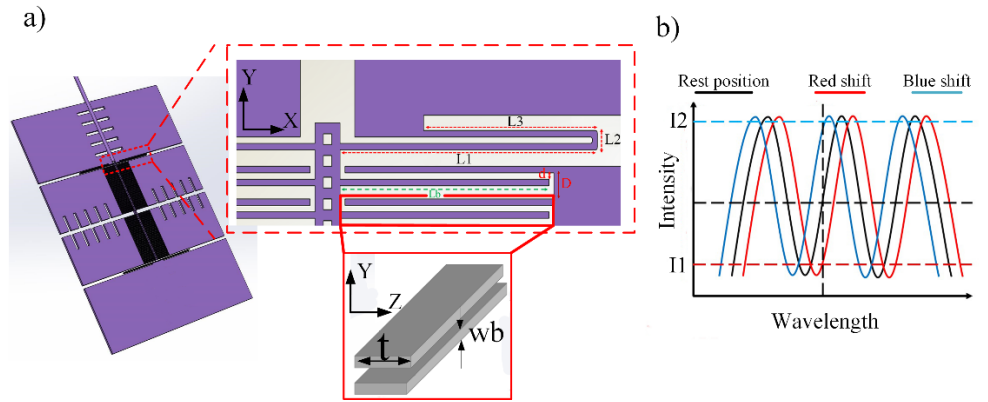

Fig. 1. a) The structure of the device. A rectangular PM is suspended by four U-shaped springs. The arrays of the movable and fixed comb fingers form flat capacitors along the sensing axis. b) The optical spectrum of the FP cavity. A spectrum shift appears when the length of the cavity changes, leading to intensity variations for a specific wavelength.

## 2.2 Optical Design

The optical system of the proposed closed-loop MOEMS accelerometer is based on the FP interferometer. The reflective surfaces of the FP cavity comprise the end of a cleaved single-mode optical fiber and the cross-section of the PM. The propagation of light in the FP cavity is simulated using the FDTD method with the parameters mentioned in Table 1. The intensity changes versus the cavity length for a laser beam with a wavelength of 1550 nm are shown in Fig. 2-a. As can be seen, the intensity variations are a periodic function of the cavity length. As a result, in the open-loop mode of the sensor, the maximum detectable mechanical displacement is limited to the region between two successive minimum and maximum intensities. Furthermore, considering the linear response range of the sensor, the maximum mechanical displacement is limited to the region highlighted in Fig. 2-a. The length of the acceptable mechanical displacement in the designed structure is 260 nm. This mechanical displacement leads to a spectral shift of 6.8 nm in the FP cavity spectrum. If the desired measurement range is ±1 g (g is the gravitational acceleration), the corresponding mechanical displacement should be limited to ±130 nm in order to obtain the linear response of the sensor. In this case, the mechanical and optical sensitivities of the sensor are 130 nm/g and 3.4 nm/g, respectively.

Table 1. The parameters used in the optical design and simulation.

| Description | Symbol | Value |
|---|---|---|
| Cavity length | $D_0$ | 42 µm |
| Source wavelength | $\lambda$ | 1.55 µm |
| Optical fiber core diameter | a | 8.2 µm |
| Optical fiber core refractive index | $n_{co}$ | 1.446 |
| Refractive index of the optical fiber cladding | $n_{cl}$ | 1.441 |
| Refractive index of the silicon | $n_{si}$ | 3.48 |
| Refractive index of the cavity medium | n | 1 |

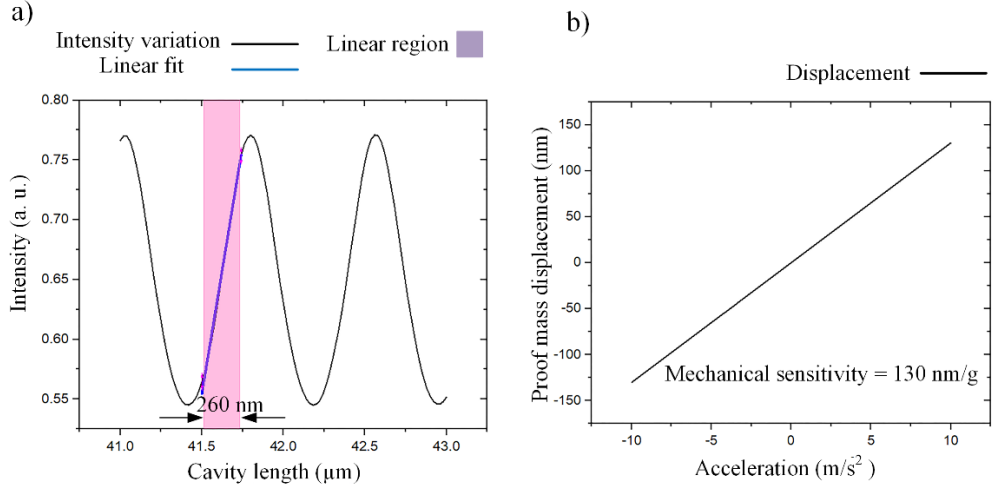

Fig. 2. a) The FP cavity output for a wavelength of 1550 nm versus the cavity length changes simulated using FDTD method. The linear response range of the FP cavity is limited to the highlighted region; b) The PM displacement versus the applied acceleration simulated in Comsol Multiphysics.

In open-loop FP-based MOEMS accelerometers, the measurement range of the sensor corresponds to the linear response range of the spectrum. Since the linear response range is limited, increasing the measurement range of the sensor decreases the sensitivity. Because there is no mechanical displacement in the closed-loop MOEMS accelerometer, this limitation will no longer be a problem. This characteristic makes the closed-loop MOEMS accelerometer an excellent sensor with a high sensitivity and a wide measurement range simultaneously.

### 2.3 Mechanical Design

The mechanical structure of the accelerometer consists of a PM which is suspended by four U-shaped springs (Fig. 1-a). This structure is designed in a way that it can move freely in the Y direction. Because of the high spring constant along the X and Z axes, the displacement along these two axes is negligible. Hence, this structure can be considered as a one-dimensional mass-spring system and described as follows [33]:

$$\ddot{y} + \gamma \dot{y} + \omega_0^2 y = -a_y, \qquad \omega_0 = \sqrt{k_y/m} \qquad (1)$$

In this equation, $m$ is the mass of the PM, $y$ is the sensing axis of the sensor, $\gamma$ is the damping coefficient, $a_y$ is the acceleration along the Y axis, $k_y$ is the spring constant of the structure along the Y axis, and $\omega_0$ is the natural frequency. The mechanical simulation of the structure is performed in the solid mechanics module of the COMSOL Multiphysics. The parameters used in the simulation are mentioned in Table 2. For a better illustration, the sizes of the springs and comb fingers are shown in Fig. 1-a. The mechanical displacement of the structure versus the applied acceleration is shown in Fig. 2-b. The mechanical displacement in the Y axis is ±132.4 nm for ±1 g which matches the desired displacement in the linear response range of the sensor. The first four resonant modes of the structure are calculated by the FEM simulation and are shown in Fig. 3. The first resonance frequency is 1403 Hz which determines the bandwidth of the sensor. This mode is completely in the sensing direction (the Y axis). The second resonant mode is at 6741 Hz and oscillates along the Z axis. The third and fourth modes are rotational modes at 7920 Hz and 8878 Hz, respectively. The second, third, and fourth frequencies are higher than the first one and do not have any effect on the performance of the sensor since, this sensor will be used at frequencies less than one-fifth the first resonant frequency mode (about 280 Hz). The results of the mechanical simulation are summarized in Table 3.

Table 2. The parameters used in the mechanical simulation.

| Parameters | Symbol | Value |
|---|---|---|
| Density of silicon | $\rho$ | 2.329 g/cm$^3$ |
| Young's modulus | E | 169 GPa |
| Poisson's ratio | $\upsilon$ | 0.28 |
| Thickness | t | 75 µm |
| Dimensions of the PM | L×W | 3580×60 µm$^2$ |
| Dimensions of the etch holes | l×w | 20×20 µm$^2$ |
| Width of the springs | $W_s$ | 10 µm |
| Length of the spring along the X axis | L1, L2, L3 | 1345, 35, 1145 µm |
| Width of the beams | $W_b$ | 12 µm |
| Length of the beams | $L_b$ | 500 µm |
| Distance between the fixed and movable beams | d | 10 µm |
| Distance between the fixed beams | D | 42 µm |

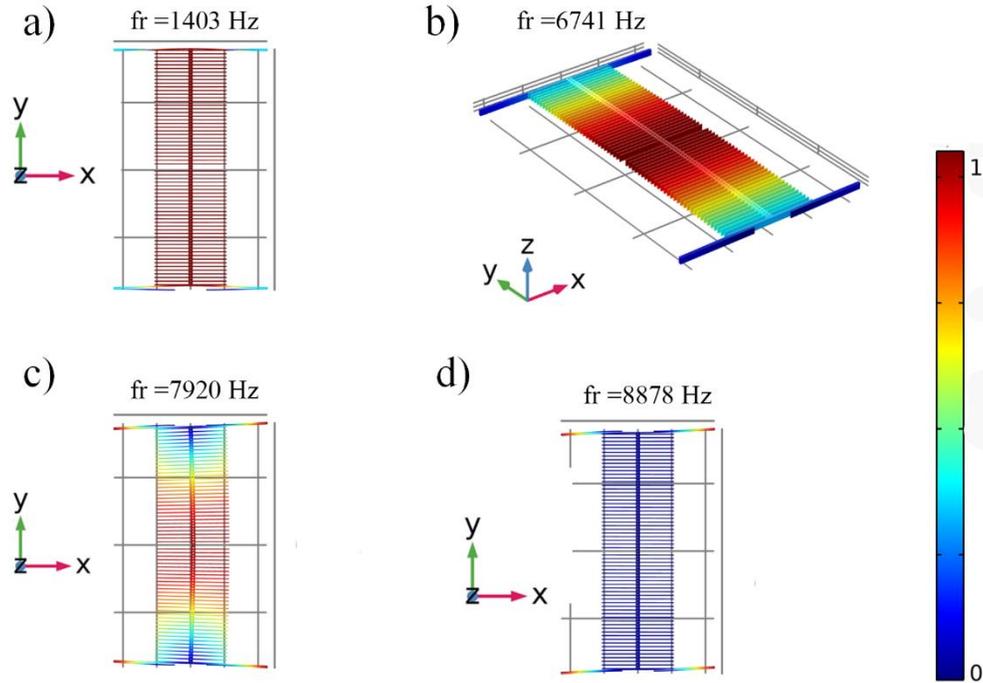

Fig. 3. The first four resonant modes. a) the first resonant mode is completely in the sensing axis of the sensor (the Y axis); b) the second mode oscillates along the Z axis; c, d) the third and fourth modes are rotational modes.

Table 3. The results of the mechanical simulation.

| Characteristic | Symbol | Simulation result |
| --- | --- | --- |
| Mass of the PM | m | 0.177 mg |
| Spring constant along the Y axis | Ky | 13.368 N/m |
| Spring constant along the X axis | Kx | $3.161 \times 10^4$ N/m |
| Spring constant along the Z axis | Kz | $1.770 \times 10^4$ N/m |
| Mechanical sensitivity along the Y axis | Sy | 132.4 nm/g |
| Mechanical sensitivity along the X axis | Sx | $5.6 \times 10^{-4}$ nm/g |
| Mechanical sensitivity along the Z axis | Sz | $1.0 \times 10^{-3}$ nm/g |

### 2.4 Electrostatic Design

As mentioned in the previous sections, in the closed-loop MOEMS accelerometer, an electrostatic force compensates the force caused by the acceleration. When an acceleration is applied to the structure, the PM tends to move in the opposite direction. However, the electrostatic force fixes the PM in its resting position. The magnitude of the electrostatic force should be the same as the acceleration force. In addition, it should be in the opposite direction. In the closed-loop MOEMS accelerometer, this electrostatic force is supplied by the array of fixed and movable comb fingers. The comb fingers act as capacitor surfaces. By applying voltage to fixed and movable fingers desired electrostatic force can be produced. The structure of one of the capacitors is schematically shown in the inset of Fig. 1-a. The arrangement of the fixed and movable comb fingers in the upper and lower parts of the structure is shown in the inset of Fig. 4-a. Two electrostatic force is imposed on each comb fingers. The separation of each successive comb fingers is shown with $d_1$ and $d_2$. The value of $d_1$ and $d_2$ in the upper part are 10 microns and 20 microns respectively, while they have reverse order in the lower part. Therefore, the total electrostatic force in the upper part is upward, whereas it is downward in the lower part. The electrostatic force between capacitor surfaces can be considered as follows [33]:

$$C = \varepsilon \frac{tl}{d} \tag{2}$$

$$F_N = \frac{1}{2}\frac{\partial C}{\partial d}V^2 = -\frac{1}{2}\varepsilon\frac{lt}{d^2}V^2 \tag{3}$$

In this equation, C is the capacitance of the two parallel surfaces, V is the applied voltage, d indicates the distance between the surfaces, ε is the permittivity, *l* and *t* stand for the length and thickness of the comb fingers, respectively. The displacement of the comb finger is considerably smaller than the separation between the capacitors' surfaces, so the variation of capacitance is negligible and the electrostatic force is proportional to $V^2$. The electrostatic force has a non-linear relationship with the voltage, while the spring elastic force has a linear relationship with the displacement. To solve this problem and create a linear relationship between the voltage and the electrostatic force, the differential method is used. As shown in Fig. 4-a, in this method, the upper and lower arrays of capacitors and the PM are connected to different voltages. Furthermore, the distance between the fixed and movable comb fingers is asymmetric in the upper and lower parts of the structure. As a result, the upper array of capacitors imposes an upward total force to the PM, whereas the lower array of capacitors imposes a downward total force to the PM. The net electrostatic force imposed on the PM can be considered as a vector summation of the upward and downward forces.

$$F_{up} \propto (V_{pm} - V_{up})^2 \tag{4}$$

$$F_{down} \propto (V_{pm} - V_{down})^2 \tag{5}$$

In these equations, $V_{pM}$ is the voltage of the PM, and $V_{up}$ and $V_{down}$ are the voltages of the upper and lower arrays of comb fingers, respectively. By setting $V_{down} = 0$, the net electrostatic force is:

$$F_{net} = F_{up} - F_{down} \propto (V_{up}^2 - 2V_{up}V_{pm}) \tag{6}$$

If $V_{up}$ be a constant voltage, there is a linear relationship between the voltage of the PM and the electrostatic force. The equation 6 shows that if $V_{PM} = V_{up}/2$, the net electrostatic force is zero. By increasing the PM voltage from this value, the total electrostatic force is downward and vice versa.

To verify this method, the displacement of the PM due to the electrostatic force is simulated in COMSOL Multiphysics. At first, the non-linear scenario is tested. The voltages of the PM and the lower electrodes are considered to be zero. The upper electrodes are connected to different voltages ranging from 0 to 10 V. The PM displacement is shown in Fig. 4-c. In this scenario, the electrostatic force has a non-linear relationship with the applied voltage ($F_N \propto V^2$). In the differential scenario, the upper electrodes are connected to a constant voltage of 10 V, whereas the lower electrodes are connected to a zero voltage. The voltage of the PM varies from 0 to 10 V. The results of the differential method are shown in Fig. 4-d. It can be seen that the displacement of the PM has a linear relationship with the applied voltage and the mechanical sensitivity of the structure is 127.3 nm/V. Hence, the sweeping voltage ranging from 0 to 10 V can move the PM ±636.5 nm from its resting position.

As mentioned in the optical design section, the maximum mechanical displacement of the PM which corresponds to the linear response range of the sensor is ±130 nm. The simulation shows that the electrostatic force can compensate a mechanical displacement of ±636.5 nm. The designed sensor has a linear response in the range of ±1 g in the open-loop mode. As a result, its measurement range can be extended to $\frac{636.5}{132.4} \approx \pm 4.8$ g in the closed-loop mode.

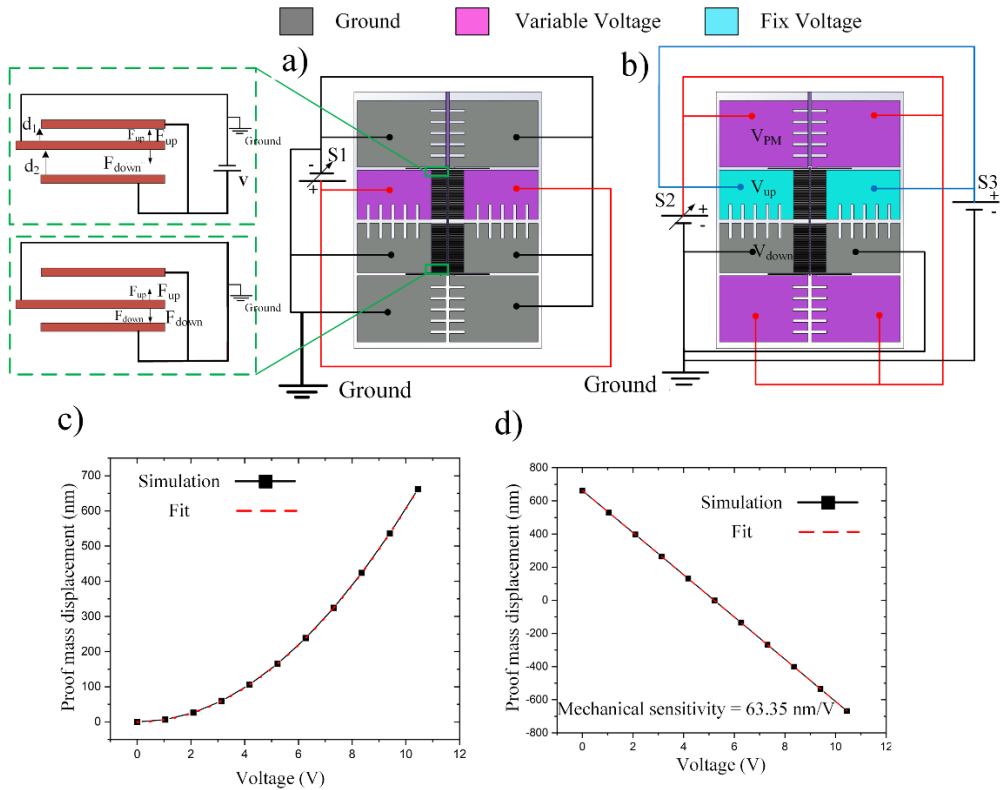

Fig. 4. a,b) The configuration of the upper and lower comb fingers producing the electrostatic force and the voltages applied to the electrodes in the non-linear mode and linear modes, respectively; c, d) the PM displacement in the non-linear and linear modes, respectively.

For the closed-loop system, a PI controller is implemented in the microcontroller (MC). The error signal of the control system indicates the displacement of the PM from its resting position which is detected by a photodiode. The photodiode is biased in the reverse condition and a transimpedance amplifier converts the photodiode current into voltage. This voltage is applied to the MC by an analog-to-digital converter (ADC). After processing the error signal, a digital-to-analog converter (DAC) is used to apply the analog voltage to the PM. This voltage is also considered as the output of the accelerometer. The transfer function of the system is calculated by Eq. 1. The damping coefficient ($\gamma$) in this equation is determined by the transient response of the PM to an electrostatic stimulus (see section 3.3). This transfer function is used in MATLAB software to determine the proper P and I coefficients for the closed-loop system. The block diagram of the closed-loop system is shown in Fig. 5.

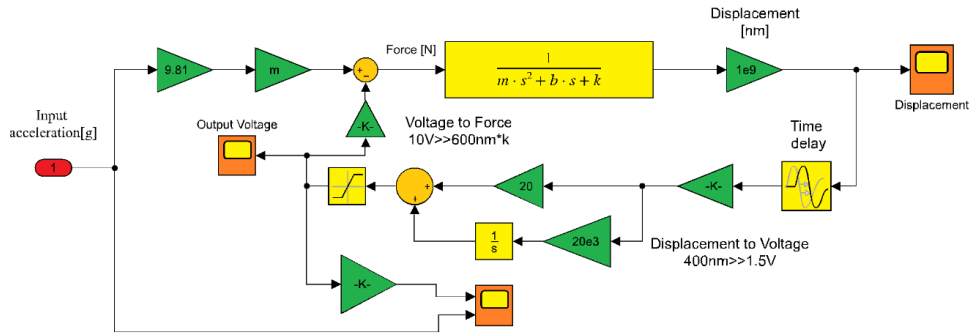

Fig. 5. The block diagram of the closed-loop system.

The desired frequency measurement range for this work is lower than 100 Hz. The frequency response of the sensor in the open-loop and closed-loop modes are shown in Fig. 6. This figure shows that in the closed-loop mode, the frequency response is enhanced considerably in comparison with open-loop mode. By increasing the gain of the PI controller, the frequency response become uniform. However, an inevitable time-delay in the control system makes the closed-loop system unstable for the higher gains. For higher gains, a few microseconds delay cause the system become unstable. To keep the system stable, the gain of the feedback loop is set to P=10 and I=10e3.

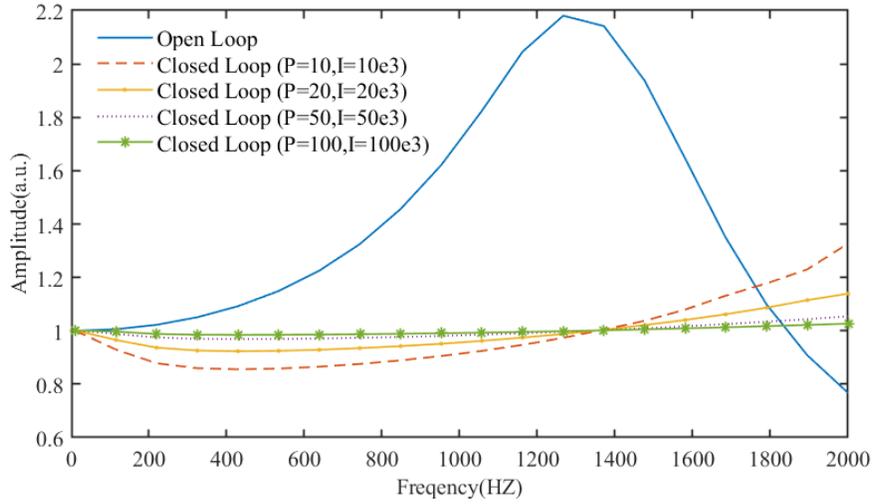

Fig. 6. Frequency response of the sensor in the open-loop and the closed-loop modes.

## 3. Fabrication and Characterization

### 3.1 Fabrication Process

The sensor is fabricated on a silicon-on-insulator (SOI) wafer using the deep reactive ion etching (DRIE) method. The wafer has a 75-micron device layer, a 4-micron oxide layer ($SiO_2$), and a 470-micron handle layer. The steps of the fabrication process are shown schematically in Fig. 7-a. At first, the SOI wafer is cleaned using the RCA procedure. Then, a layer of AZ photoresist with a thickness of 1 micron is spin-coated. The pattern of the sensor is transferred to the photoresist using contact-mask lithography. After developing, the AZ mask is created on the wafer which is ready for the DRIE process. The structure of the sensor consists of the PM and springs formed in the DRIE process. Then, the oxide layer is removed using HF vapor and the PM is released and moves freely. After the lithography process, a thick layer of gold is coated on the edge of the PM using the sputtering layer deposition method. This gold layer increases the reflectivity of the PM cross-section as a surface of the FP cavity and prevents unwanted reflection from the inside of the PM. Then, wire bonding is used for the electrical connections.

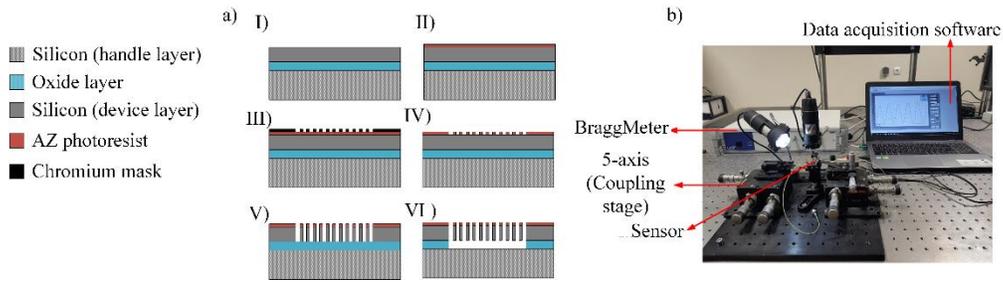

Fig. 7. a) The fabrication process including six main steps including (I) substrate cleaning (RCA procedure), (II) photoresist coating, (III) exposing the masked photoresist to UV light, (IV) photoresist development, (V) DRIE process, and (VI) lifting-off the sacrificial layer; b) the optical fiber assembly setup in the sensor.

After the fabrication process, a cleaved single-mode optical fiber is placed between the designed grooves in a 5-axis coupling stage. The FP cavity is formed between the end of the optical fiber and the cross-section of the PM. The other end of the optical fiber is connected to a BraggMETER (FS22 Industrial BraggMETER) which sends the optical signal to the cavity with a wavelength of 1500-1600 nm (Fig. 7-b). The length of the cavity is adjusted according to the number of peaks in the optical spectrum of the cavity. In the design section, the length of the cavity was selected 40 microns in accordance with about 3 peaks in the optical spectrum of the BraggMETER. Finally, the optical fiber is fixed in its place using a UV-cured adhesive (Norland 61). The SEM image of the fabricated device is shown in Fig. 8.

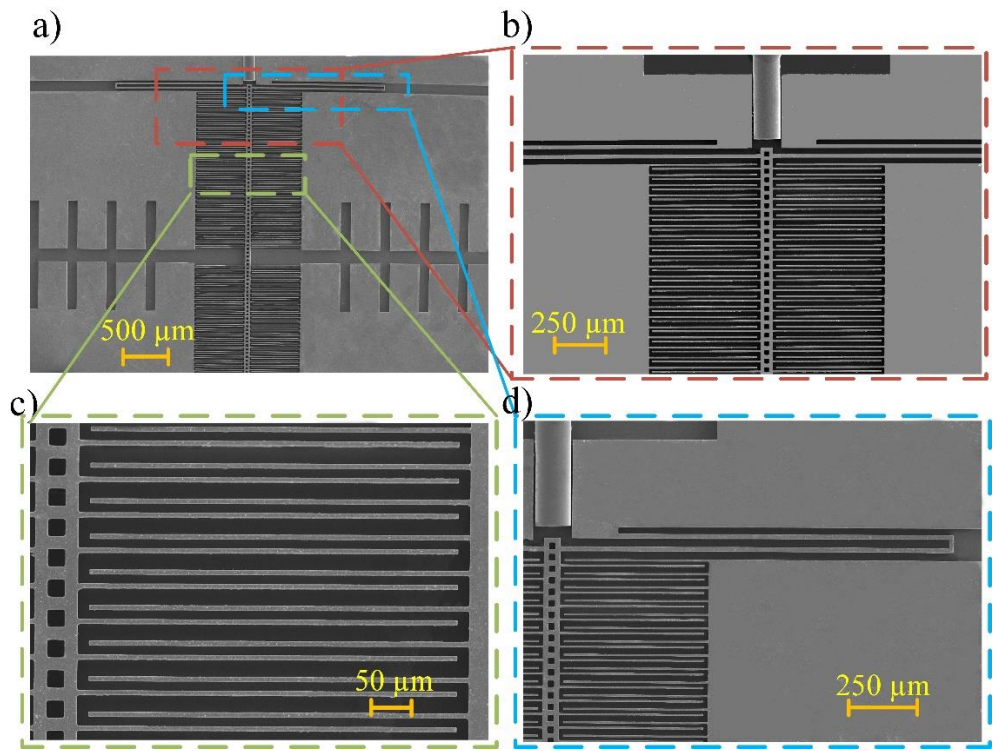

Fig. 8. The SEM image of the fabricated device.

## 3.2 Open-Loop Characterization

In the first step of characterization, the differential method which produces a linear electrostatic force is used and its results are tested. The optical setup (Fig. 7-b) consists of a BraggMETER, a sensor, and an electrical power supply. The output fiber of the sensor is connected to the BraggMETER and the output spectrum of the FP cavity can be monitored using a PC. To evaluate the response of the sensor, the same voltages used in the simulation section are applied (Fig. 4-a and b). The results of applying voltage in the non-linear and linear modes (the differential method) are shown in Fig. 9-a and Fig. 9-b, respectively. They show that the response of the sensor to the electrostatic force is linear when the differential method is used. The FP cavity length is calculated as follows [34]:

$$L = \frac{\lambda^2}{2\Delta\lambda_{FSR}} \tag{7}$$

In this equation, $\Delta\lambda_{FSR}$ is the spacing in wavelength between two successive peaks (deeps) and $\lambda$ is the geometric means of the wavelengths at those peaks (deeps). Therefore, the mechanical displacement of the PM is calculated based on the FP cavity length before and after the displacement. Fig. 9-c shows that the mechanical sensitivity is 124.3 nm/V which is in good agreement with the simulation results. Therefore, the electrostatic force can compensate the 1243-nm displacement of the PM due to the applied acceleration. The maximum mechanical displacement of the PM in the open-loop mode is 260 nm, while the measurement range in the closed-loop mode can be extended to the ratio of $\frac{1243}{260} \approx 4.8$.

To evaluate the optical sensitivity of the sensor, it is placed on top of a turntable. The rotation of the turntable can provide a variable acceleration ($a = g\sin\theta$). The optical response of the sensor to the static acceleration in the open-loop mode is shown in Fig. 9-d. The optical sensitivity of the sensor is 3.4 nm/g.

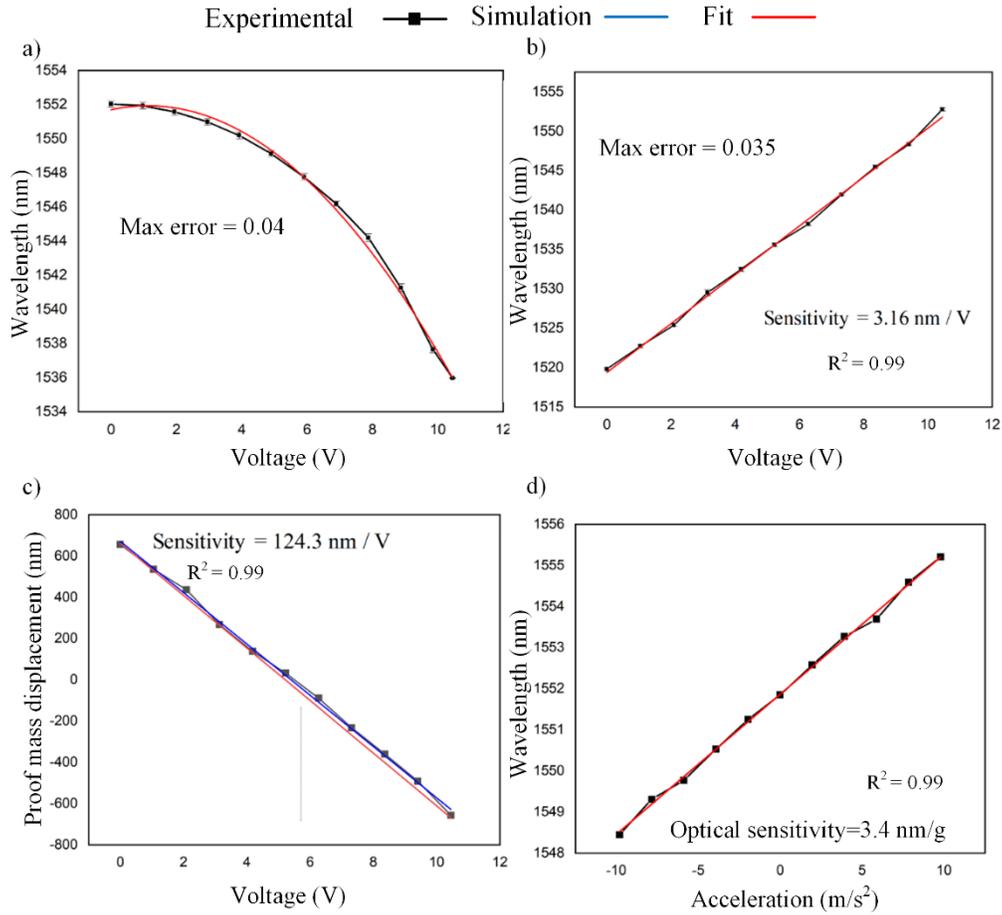

Fig. 9. a) The FP spectrum shift due to the applied voltage in the non-linear mode; b) the FP spectrum shift due to the applied voltage in the linear mode (the differential method); c) the PM displacement due to the applied voltage ranging from 0 to 10 V in the linear mode; d) the sensor response to the acceleration in the range of ±1 g in the open-loop mode.

### 3.3 Closed-Loop Characterization

In this step, the accelerometer is characterized in the closed-loop mode. The test setup consists of a sensor, a tunable laser source (Thorlabs TLX1 C-Band Tunable Laser), an isolator, a photodiode (Thorlabs-PDA10CS-EC), a turntable, a commercial accelerometer (ADXL1002), and a control system. The schematic of the experimental setup is shown in Fig. 10-a. The output of the source passes the isolator and is sent to the FP cavity. The reflected light is sent to the photodiode through a coupler. In this setup, the commercial accelerometer is used to measure the applied acceleration. The voltage applied to the comb fingers is measured and reported as the output signal of the sensor. The results of applying a static acceleration (±1 g) to the closed-loop sensor are shown in Fig. 10-b. The sensitivity of the sensor in this mode is 1.16 V/g. In the closed-loop mode, the PM ideally does not have any movement, therefore, the output signal of the photodiode should remain fixed. Fig. 10-b shows that the photodiode voltage has an approximately constant value in the measurement range.

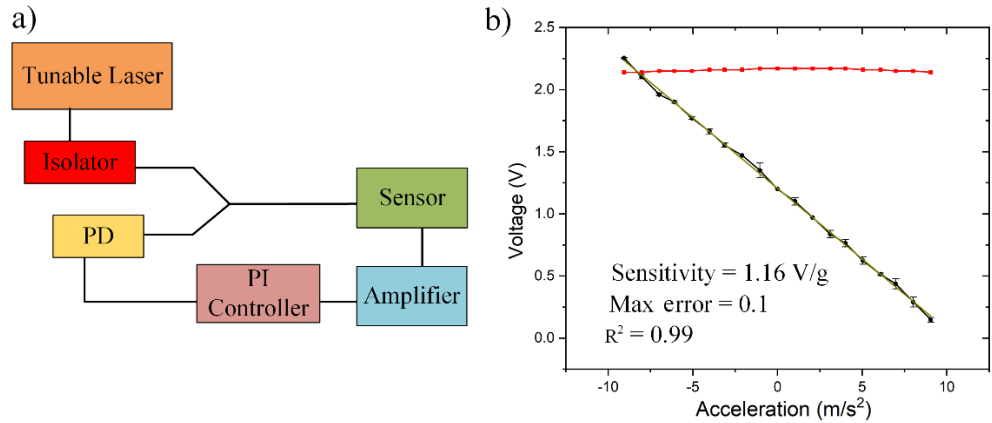

Fig. 10. (a) The schematic of the experimental setup for the characterization of the closed-loop mode; b) the static response of the sensor in the closed-loop mode. In this mode, the photodiode voltage reamins approximately fixed (the red values) and the applied voltage to the comb fingers is the sensor output (the black values).

To investigate the dynamic behavior of the sensor, the turntable is replaced by a shaker. The shaker can apply a horizontally sinusoidal acceleration. With a frequency of 100 Hz, an amplitude of 1-5 g, and 1-g steps, the dynamic acceleration is applied to the sensor. The output voltage of the sensor is shown in Fig. 11-a. Fig. 11-b shows the responses of the sensor and the reference accelerometer when a 5-g acceleration with a frequency of 100 Hz is applied. As shown in Fig. 11-c, the output of the sensor is linear in the measurement range of ±5 g and the output of the photodetector remains constant. Moreover, the Allan deviation analysis is performed to calculate the bias stability of the closed-loop accelerometer. As shown in Fig. 11-d, the bias instability of the sensor is 40 µg.

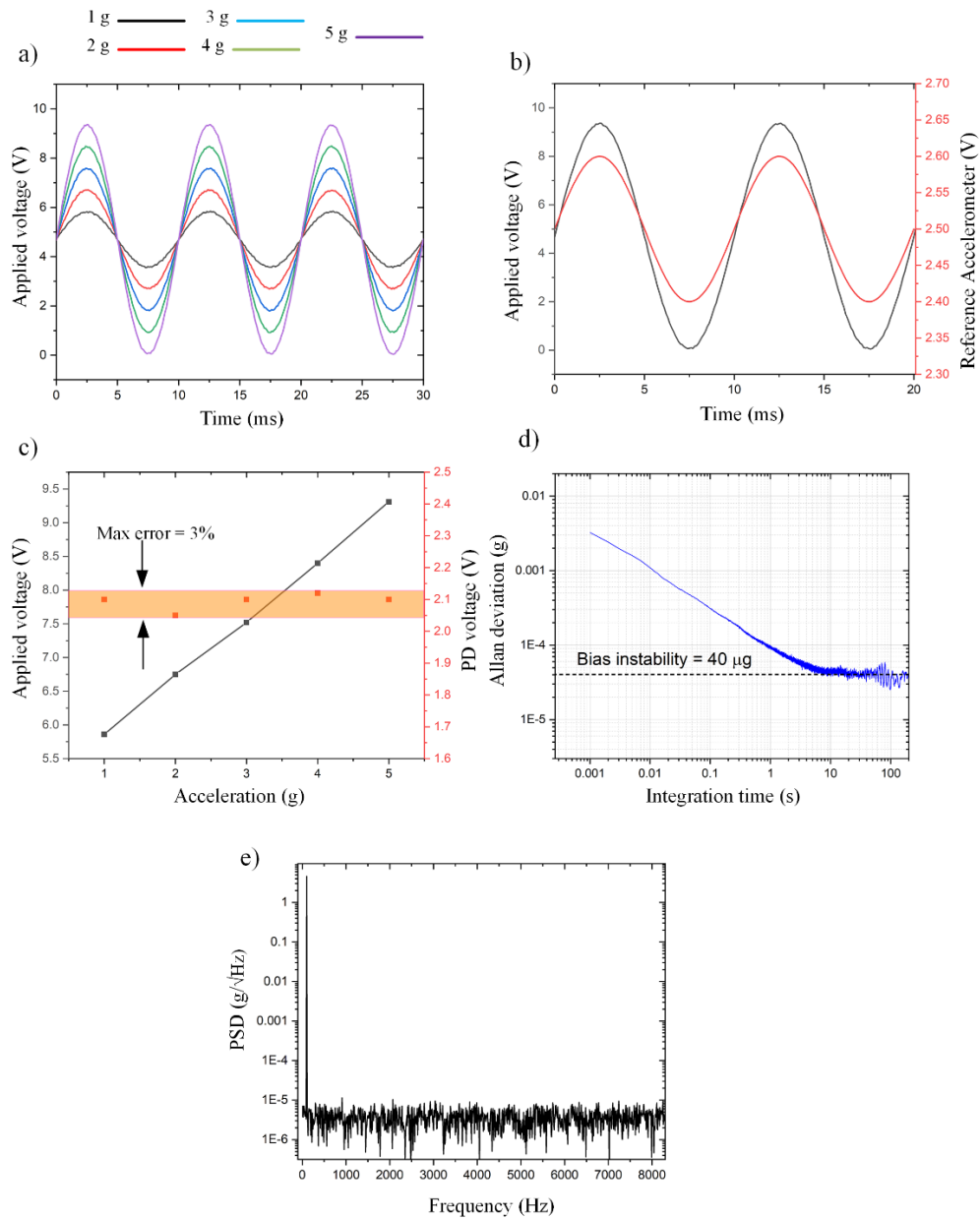

Fig. 11. a) The dynamic response of the sensor to five different accelerations with the same frequency of 100 Hz; b) the response of the sensor to the acceleration of 5g in comparison with the reference accelerometer, c) linearity of the applied voltage to PM as a respons of the sensor (black points and left axis) and stability of the PD output voltage (red points and right axis), d) Allan deviation of the test data, e) The PSD function of the sensor output in 5g sinusoidal acceleration.

To investigate the damping behavior of the sensor, a step voltage with a frequency of 100 Hz is applied to the comb fingers. After each pulse, the structure acts as a damped harmonic oscillator. The response of the sensor to the step voltage is shown in Fig. 12. By fitting an exponential curve, the damping coefficient ($\gamma$) of the mechanical structure is obtained. Also, the quality factor of the mechanical system (Q) can be calculated by the following equation using the resonance frequency of the vanishing oscillations based on experimental data (1280 Hz) shown in Fig. 12 [35].

$$Q = \frac{\sqrt{\omega_0^2 - \gamma^2}}{2\gamma} \tag{8}$$

Based on the experimental data, for this mechanical structure γ and Q are 694 s$^{-1}$ and 5.8, respectively. Furthermore, the rise time of the sensor is 182 µs as shown in Fig. 12.

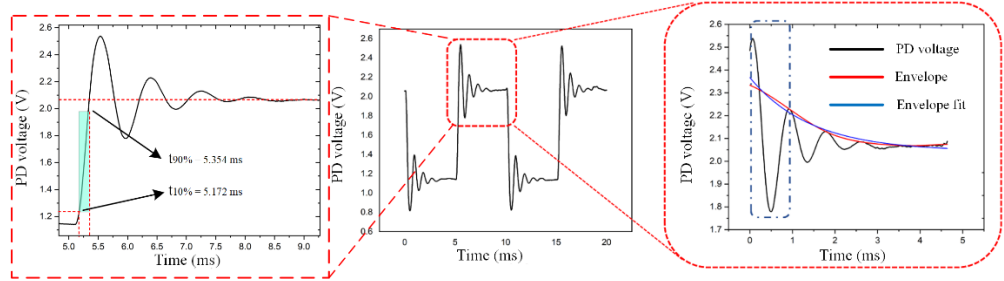

Fig. 12. The response of the sensor to the step volatge and damped oscillation after each pulse.

## 4. Comparison and Analysis

To compare the performance of the proposed sensor with those of other sensors in this field, the main characteristics of the proposed sensor and three recent accelerometers are summarized in Table 4. The closed-loop MOEMS accelerometer has a wider measurement range and a higher sensitivity than the open-loop FP-based accelerometer in our previous study[21]. The measurement range of the proposed sensor is five times greater than that of the accelerometer in [21] and its bias instability is reduced to 22.8 µg. Furthermore, the dimensions of the closed-loop MOEMS accelerometer are in the millimeter scale and its structure is completely integrated, whereas the closed-loop micro-grating accelerometers in [36,37] are bulky and have higher volumes and masses. Moreover, the fabrication process of the proposed sensor is based on a straightforward bulk micromachining method which facilitates the mass production of the device.

Table 4. The comparison of the main characteristics of the closed-loop MOEMS accelerometer with those of the accelerometers in three recent works.

| Characteristic | This work | 2021[21] | 2021 [37] | 2019 [36] |
|---|---|---|---|---|
| Measurement range | ±5 g | ±1 g | ±1 g | ±1 g |
| Sensitivity | 1.6 V/g | 550 mV/g | - | $6.53 \times 10^3$/g |
| Bias stability | 25.8 µg | 310 µg | 1.3 µg | - |
| Resonance frequency | 1403 Hz | 1382 Hz | 218.4 Hz | - |
| Bandwidth | 500 Hz | 300 Hz | - | 525 Hz |
| Non-linearity | | 0.01 | 0.35% | 0.28% |
| Mass of PM | 0.177 mg | 0.94 mg | 1.78 g | - |
| Sensing structure | FP micro-cavity (Closed-loop) | FP micro-cavity (Open-loop) | Micro-grating (Closed-loop) | Micro-grating (Closed-loop) |

## 5. Conclusion

In this paper, the design and fabrication of a closed-loop MOEMS accelerometer was reported. The working principle of the designed accelerometer was based on a FP interferometer and a closed-loop detection method. The proposed structure consisted of a PM and comb finger arrays. The electrostatic force between the fixed and movable comb figures compensated the movements of the PM due to the applied acceleration. The voltage required to produce this electrostatic force was considered as the output of the sensor. The closed-loop MOEMS accelerometer has a wider measurement range and a higher sensitivity than the open-loop FP-based accelerometer in our previous study [22]. The measurement range of the proposed sensor is five times greater than that of the accelerometer in [22] and its bias stability is reduced to 40 µg. Furthermore, the dimensions of the closed-loop MOEMS accelerometer are in the millimeter scale and its structure is completely integrated. Moreover, the fabrication process of the proposed sensor is based on a straightforward bulk micromachining method which facilitates the mass production of the device. The characterization of the sensor showed that it could measure acceleration in the range of -5 g to +5 g where the sensitivity of the sensor was 1.16 V/g.

**Disclosures.** The authors declare no conflicts of interest.

**Data availability.** Data underlying the results presented in this paper are not publicly available at this time but may be obtained from the authors upon reasonable request.